\documentclass[prd,preprint,tightenlines,floatfix,
showpacs,preprintnumbers,nofootinbib,eqsecnum,superscriptaddress]{revtex4}

\usepackage[T1]{fontenc}		

\usepackage{amsmath,amsfonts,amssymb,amstext,mathrsfs}
\usepackage{mathpazo}

\usepackage[dvips]{graphicx}
\usepackage{epsf,float}
\usepackage{revsymb}

\usepackage{dcolumn}
\usepackage{braket}
\usepackage{color,xcolor}
\usepackage{graphicx}
\usepackage{subfigure}
\usepackage{multirow}
\usepackage{tabularx}
\usepackage{pstricks}
\usepackage[section]{placeins}
\usepackage{booktabs}
\usepackage{array}

\usepackage{hyperref}

\begin{document}

\begin{center}

{\bf\Large
On the Implications of Energy and Momentum Conservation for Particle Emission in A+A Collisions at SPS Energies
}

\vspace{0.5cm}

Antoni Szczurek$^{1,2}$, 
Miros\l{}aw Kie\l{}bowicz$^1$, 
Andrzej Rybicki$^1$\\

\vspace{0.6cm}

{\small $^1$~H.~Niewodnicza\'{n}ski Institute of Nuclear Physics, Polish 
 Academy of Sciences, Radzikowskiego 152, 31-342~Krak\'ow, Poland\\
 $^2$~University of Rzesz\'ow, Rejtana 16, 35-959 Rzesz\'ow, Poland}
\end {center}

\vspace*{0.0cm}
{\small\bf Abstract}\\

{\small
{We construct a simple model of heavy ion collisions, local in the impact parameter plane, and appropriate for the SPS
energy range. This model can be regarded as a new realization of the ``fire-streak'' approach, originally applied to studies
of lower energy nucleus-nucleus reactions.}
Starting from local energy and momentum conservation, we
nicely describe the broadening of the pion rapidity distribution
when going from central to peripheral Pb+Pb collisions {at $\sqrt{s_{NN}}=17.3$~GeV.}
The results of our calculations are compared with SPS experimental data. 
  We discuss the resulting implications on the role of energy and momentum conservation for the dynamics of particle production in heavy ion collisions. 
%
%
A specific space-time picture emerges, where the longitudinal evolution of the system strongly depends on the position in the impact parameter (${b_x}$, ${b_y}$) plane. 
%
%
This picture is consistent with our earlier findings on the longitudinal evolution of the system as deduced from electromagnetic effects on charged pion directed flow, and can provide an explanation for specific low-$p_T$ phenomena seen in the fragmentation region of Pb+Pb collisions.}

\section{Introduction}
\label{intro}

While the presence of energy and momentum conservation is evident at every stage of the heavy ion collision, it is known that its detailed impact on multiparticle production phenomena can be far from obvious. In the case of complex and mostly non-perturbative phenomena underlying the bulk of particle production, this impact should be traced with due care as it constitutes a basis for all further nontrivial phenomena like quark gluon plasma formation as well as the collective expansion of the system up to the final state observed in the detector.

In the following we shall discuss this issue for the case of the space-time picture of charged pion production at SPS energies. We will 
{formulate}
a simple model with exact local energy-momentum conservation in the initial state of the collision followed by a very simple scheme of subsequent particle production. With this simple model we will explain the centrality dependence of charged pion production as a function of rapidity which we will compare to experimental data at SPS energies. We will examine the implications of energy momentum conservation for the longitudinal evolution of the presumably deconfined matter created in the collision as a function of impact parameter, where, especially in peripheral collisions, the ``hot'' system of dense partonic matter located close to mid-rapidity is accompanied by ``colder'' but still highly energetic volumes of matter positioned in different regions of the (${b_x}$, ${b_y}$) plane and moving with large longitudinal velocities. We will address the issue on how the presence of these rapidly moving 
``trails''
could explain our earlier findings on the space-time evolution of the system as deduced from electromagnetic interactions,
as well as specific low-$p_T$ phenomena present at high pion rapidities.

In our recent studies of the above-mentioned electromagnetic (EM) effects, induced by the spectator charge on the final state spectra of $\pi^+$ and $\pi^-$ mesons produced in the non central heavy ion collision~\cite{twospec07,Rybicki_v1,twospec_auau}, we found that both the EM-distortion of $\pi^+/\pi^-$ ratios as well as the charge splitting of pion directed flow, $v_1$, offer sensitivity to the space-time evolution of pion production. 
A consistent picture emerges where the distance between the pion emitted at freeze-out and the spectator system decreases with increasing pion rapidity (faster pions are produced closer to the spectator system)~\cite{wpcf}. With the present work we aim at the construction of a simple but sufficiently realistic model which will contain the above feature in a natural way. This model, not too complicated and therefore not too time-consuming in the calculation of the response of charged pions to the electromagnetic fields present in the collision, 
will be used
in our further studies of electromagnetic effects on multiparticle
production phenomena.

{After a first version of this 
paper was completed we 
learned about 
the
``fire-streak''
  model, proposed a long time ago by Myers \cite{M1978} and applied to
  nuclear collisions by Gosset, Kapusta and Wesfall \cite{GKW1978}.
We note the evident similarity of our approach to the nuclear fire-streak model.
The
 early papers \cite{M1978,GKW1978} concentrated rather on low energies, below 
the
transition
to quark-gluon plasma. In our case 
(the SPS energy range), 
it is commonly
believed that the phase transition takes place. Therefore our 
``fire-streaks'' are related to quark-gluon plasma matter in contrast to
the nuclear fire-streak model where they are related to nuclear matter.
There the interest was in production of $p$, $n$, $d$, $t$, $^3He$ or
$^4 He$ \cite{GKW1978}. There are also applications of a similar
concept of energy-momentum conservation to RHIC energies 
\cite{MCS2001,MCS2002,MK2002}.
There, the general picture was supplemented 
by
modelling nuclear transparency 
which introduces some new interesting aspects but also brings some 
additional uncertainties.
In \cite{MCS2001,MCS2002} the fire-streak 
picture was supplemented by a color-rope picture in order
to understand baryon stopping.

In the case considered by us the situation is intermediate between
the nuclear case ($\sqrt{s_{NN}} <$ 1 GeV) and the case of quark-gluon
plasma, where the 
``projectile streaks'' pass through the ``target streaks''
 ($\sqrt{s_{NN}} \sim$ 100 GeV), phenomenon known as transparency.
At the SPS energies ($\sqrt{s_{NN}} \sim$ 10 GeV) one expects almost
full stopping of the QGP matter. Then the situation is unique as almost
the whole modeling is determined by energy-momentum conservation
and there is no need for additional modeling.
On the other hand, the possibility of comparison of model results with experimental data as a function of impact parameter (centrality), was not available in the lower energy works \cite{M1978,GKW1978} but constitutes the main consistency test for the approach formulated
in this paper. In our view, this offers 
more restrictive conditions for testing the different detailed features postulated for the present version of the model, like e.g. the universality of pion emission functions 
claimed
in section~\ref{qgphadrons}. This in turn gives us better confidence in the conclusions on the space-time picture and longitudinal evolution of pion production, which we formulate in section~\ref{conclusions}.
}

\section{A simple energy-momentum conservation model for ``stopping'' of nuclear matter}
\label{simplemodel}

There is no 
precise knowledge
on initial space-time conditions of 
the QGP for hydrodynamical evolution (if it applies at given collision energy) as well as for
{the particle
freeze-out space-time moment.}
Therefore, in our considerations presented here, we 
avoid 
the
application of any 
specific
{microscopic}
model.
%
At SPS energies the QGP matter produced in central collisions 
is believed to be ``stopped'' (partons are slowed down) in the overall 
center-of-mass system, the QGP system expands thermally, cools down
and at a certain space-time moment, when its density is small enough, 
particles (mostly pions) are produced.

\begin{figure}[!ht]
\includegraphics[width=6cm]{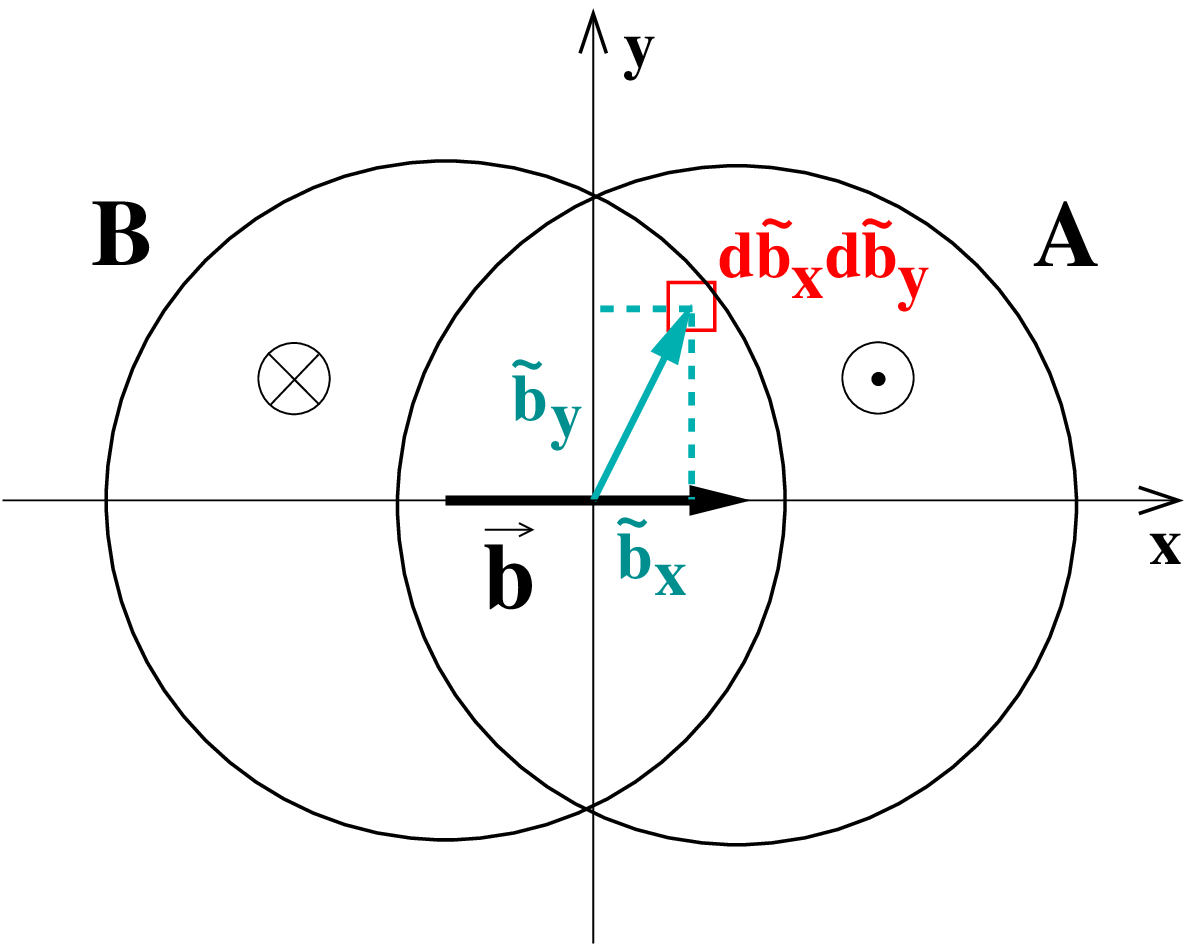}
{\caption\small 
A schematic view in the plane perpendicular to the collision axis,
assumed to be $z$. Here we introduce the local 
coordinates
$({\tilde b}_x,{\tilde b}_y)$ 
in this plane, which give the 
position of the rectangular ``strip'' ({``fire-streak''}) parallel to 
the $z$ axis as discussed in the text.
The impact parameter vector $\vec{b}=(b,0)$
is
shown in addition.
\label{bxby}
}
\end{figure}

At relativistic energies the longitudinal motion is much ``faster'' 
than the transverse motion 
at least in the first stage
of the collisions (before the transverse flow builds up).
Therefore partons can be treated separately in different ``strips''
({``fire-streaks'' \cite{M1978}}) parallel
to the collision axis (assumed to be in the $z$ direction).
The general notation is introduced in Fig.~\ref{bxby}.


\begin{figure}[!ht]
\includegraphics[width=7cm]{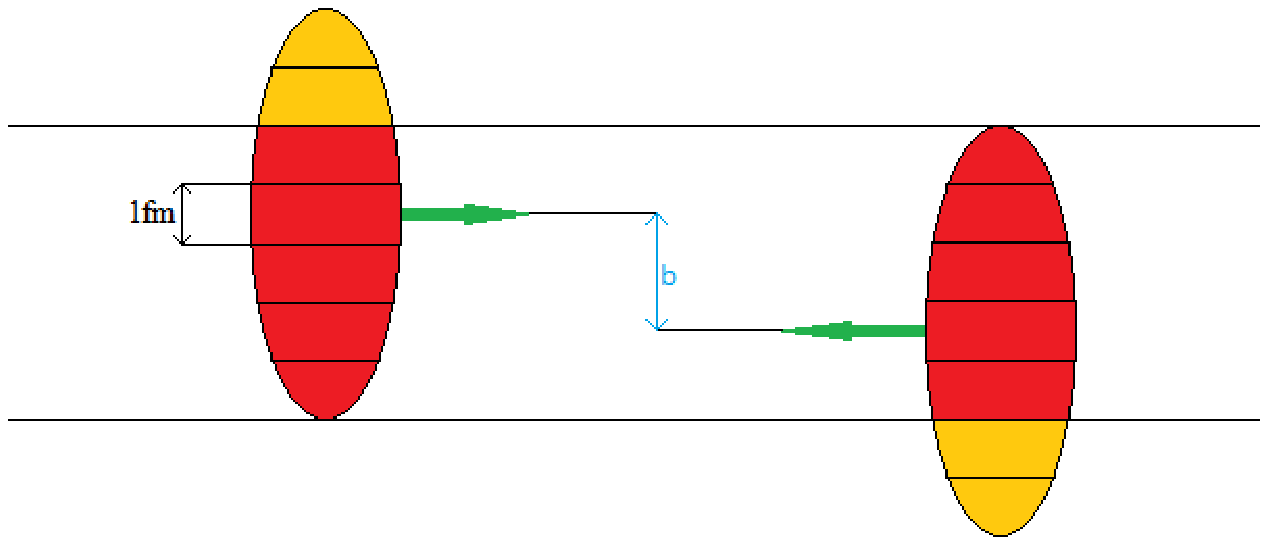}
{\caption
\small
The situation before the collision.  
For clarity, the strips 
$(d\tilde{b}_xd\tilde{b}_y)$ are shown with a finite size.
%
From the Euclidean geometry one can observe the mass difference of 
the pieces of nucleus A and nucleus B belonging to a given 
$(d\tilde{b}_xd\tilde{b}_y)$
strip.
This disbalance depends on the strip position in the transverse plane,
which can be characterized by the 2-dim $({\tilde b}_x,{\tilde b}_y)$ point.
\label{fig:ideowy_przed}
}
\end{figure}

In peripheral collisions (0 $\ll b < R_A + R_B$) the full stopping
in the overall center-of-mass system cannot take place due to 
the disbalance of masses, and in consequence also of momenta of $(d\tilde{b}_xd\tilde{b}_y)$ strips 
originating from nucleus $A$ and nucleus $B$. This is illustrated in Fig.~\ref{fig:ideowy_przed} (the
reader is invited to compare the respective opposed areas
for the selected finite size strip).

\begin{figure}[!ht]
\includegraphics[width=10cm]{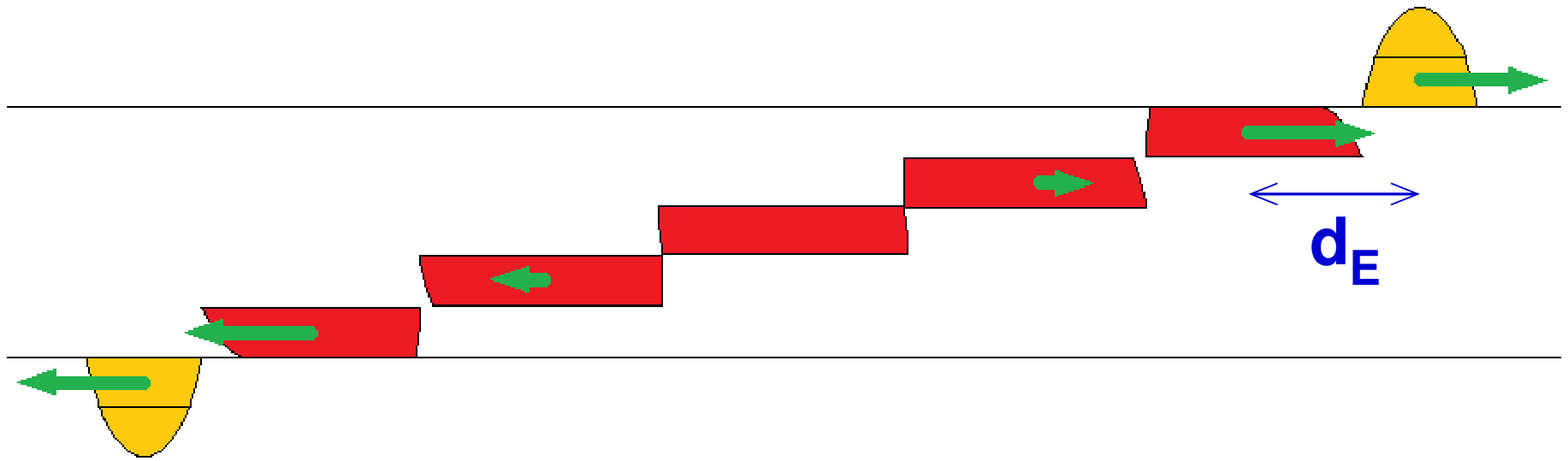}
{\caption
\small 
The situation after the collision. The area marked in red shows the partonic
matter. Each $(d {\tilde b}_x d {\tilde b}_y)$ element moves with a
different longitudinal velocity which can be obtained from
energy-momentum conservation of the matter 
contained
in the $(d {\tilde b}_x d {\tilde b}_y)$ tube.
The longitudinal ``emission distance'', 
marked $d_E$ in the plot, 
will be discussed later in the text.
\label{fig:ideowy_po}
}
\end{figure}

Then the situation is a bit more complicated.
The matter does not stop locally in 
the
$({\tilde b}_x,{\tilde b}_y)$ space.
A realistic model must include momentum conservation which leads
to final overall longitudinal velocity of the compressed strips 
of partonic matter
in the $z$
direction. The situation after the collision is illustrated in 
Fig.~\ref{fig:ideowy_po}. 

In a practical calculation one can divide the 
impact parameter
transverse plane into
two-dimensional bins ($\Delta {\tilde b}_x, \Delta {\tilde b}_y$). 
The size of the bin(s) should not be too large. 
For the work discussed in this paper we take 1~fm~$\times$~1~fm which we judge small enough for the case of Pb+Pb collisions considered here and which, on the other hand, allows a clear presentation of the results.
Then the masses originating from nucleus A and B are
calculated for each bin (strip) separately from a realistic nuclear density distribution~\cite{Trzcinska01,atph}. Subsequently we consider the 
collision of two masses $\Delta M_A$ and $\Delta M_B$ which fuse
producing a moving mass $\Delta M = \Delta M_A + \Delta M_B$.
The total momentum of the strip is $\Delta \vec{p} = \Delta \vec{p}_A +
\Delta \vec{p}_B$ and its energy is $\Delta E = \Delta E_A + \Delta
E_B$. The energy and momentum of the strip 
parts originating from the nucleus A (B) are proportional to the energy and momentum of each nucleus in the collision c.m.s., 
$\Delta E_A = E_A \cdot \Delta M_A/M_A$,  
$\Delta \vec{p}_A = \vec{p}_A \cdot \Delta M_A/M_A$, and similarly for B.  
We trivially note that for $\Delta M_A > 0$ and $\Delta M_B > 0$, 
the strip 
energy in its own c.m.s. frame, $\Delta E^*=\sqrt{\Delta E^2 - \Delta p^2}$, exceeds its mass $\Delta M$. Thus
the strip is also excited with some excitation distribution
$dE^{*}/d {\tilde b}_x d {\tilde b}_y({\tilde b}_x,{\tilde b}_y;b)$ which we approximate by 
$\Delta E^{*}/\Delta {\tilde b}_x \Delta {\tilde b}_y({\tilde b}_x,{\tilde b}_y;b)$.

We note that there is no need for us to address the exact physical
nature of the 
strips 
which are representative of the (average)
kinematical features of the matter participating in the heavy ion collision,
resulting from energy-momentum conservation. As each strip has both mass and momentum, one can
calculate the corresponding velocity, or rapidity, of the fused strips originating
from nucleus $A$ and $B$. 
This velocity 
is then 
a
function of 
the
transverse position of the strip, 
$v_S = v_S({\tilde b}_x,{\tilde b}_y)$.


\section{Pb+Pb collisions at SPS energy}
\label{pbpb}

\begin{figure}[!ht]
\includegraphics[width=7cm]{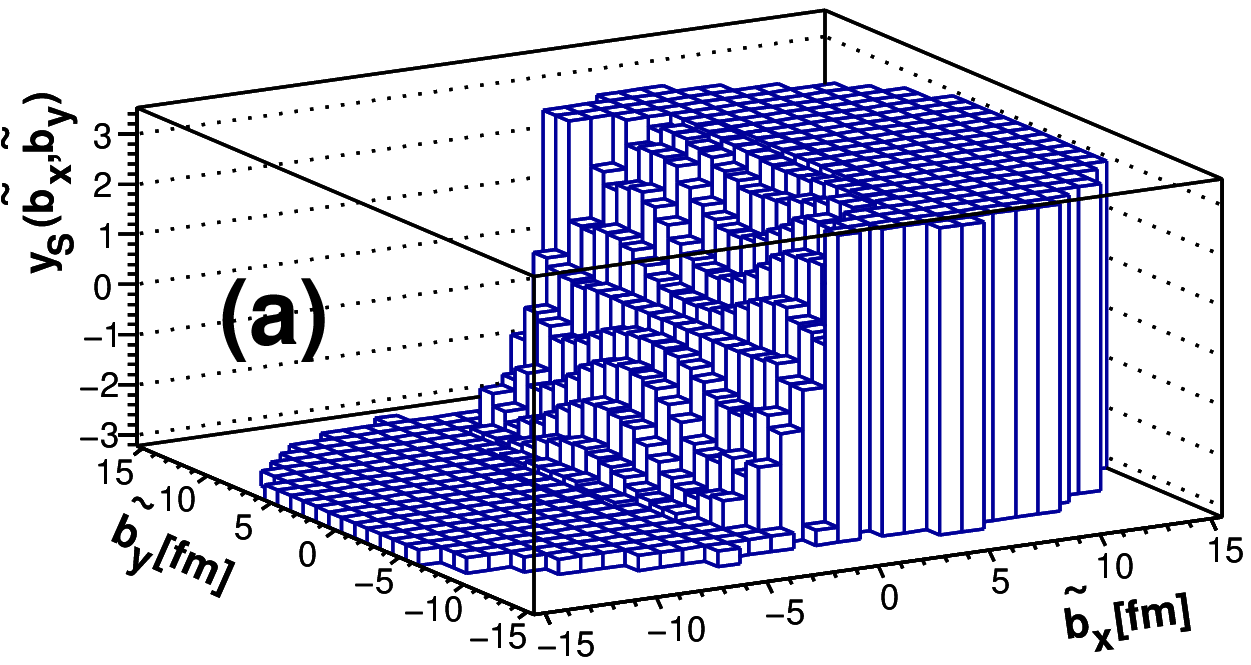}
\includegraphics[width=7cm]{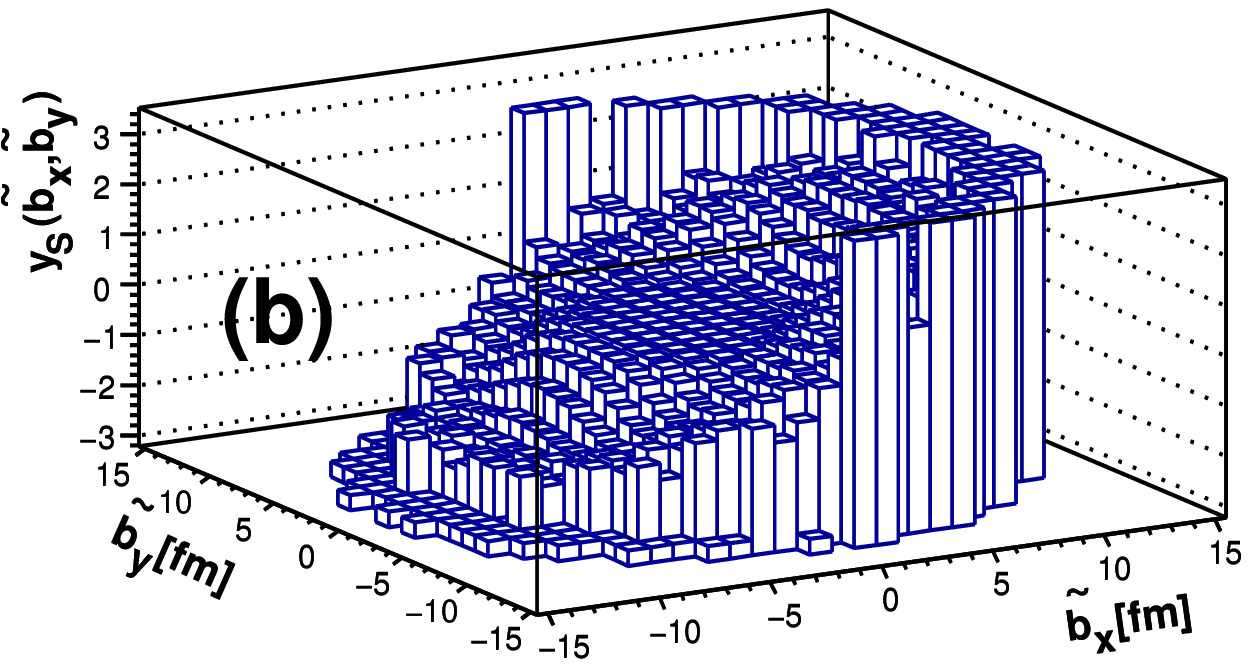}
{\caption
\small 
The $({\tilde b}_x,{\tilde b}_y)$ dependence of the matter rapidity in 
the bin (strip), for peripheral ($b$=9.72 fm, left panel) and central ($b$=2.55 fm, right panel) Pb+Pb collisions at $\sqrt{s_{NN}}=17.3$~GeV.
\label{fig:y_bxby}
}
\end{figure}

In a first step we will discuss the kinematical features of the state of matter 
initially created in Pb+Pb collisions, at top SPS energy corresponding to $\sqrt{s_{NN}}=17.3$~GeV in the collision c.m.s. This rapidity distribution of the strips of matter is illustrated in Fig.~\ref{fig:y_bxby} as a function of the position of the strip in the $({\tilde b}_x,{\tilde b}_y)$ plane perpendicular to the collision axis. The two considered values of impact parameter ($b$=9.72 fm and $b$=2.55 fm) are chosen such that they correspond to the average characteristics of ``most peripheral'' (``C4'') and ``most central'' (``C0'') centrality samples presented in the experimental paper by the NA49 Collaboration~\cite{SPS}.
As such, importantly, our simulation of ``central'' collisions corresponds roughly to realistic experimental conditions rather than to some idealized ($b=0$) case, which is not attainable experimentally.
When transforming from the average number of wounded nucleons quoted therein
to impact parameter we use a
Monte Carlo code based on the Glauber picture~\cite{Glauber} (methodology identical to that used in~\cite{twospec07}). 

Several features become immediately apparent from
Fig.~\ref{fig:y_bxby}. In peripheral collisions, the region of strip
rapidity significantly different from $\pm y_{beam}$ ($\pm$~2.9 units at
top SPS energy) corresponds to a relatively narrow range in ${\tilde
  b}_x$ (i.e., along the impact parameter vector). The overall rapidity
of the created matter depends strongly on its $x$ position and the
``central'' rapidity region ($-1<y_S<1$) remains confined to no more
than 4~fm in ${\tilde b}_x$, less than half the distance between the two
nuclear centres. Nontrivially, a weaker but also significant variation
of strip 
 rapidity is apparent in the ${\tilde b}_y$ direction. Thus, from simple energy-momentum conservation, the different sub-regions of the naive ``reaction lens'' shown in Fig.~\ref{bxby} will move away from each other, with large relative velocities.

\begin{figure}[!ht]
\hspace*{-0.2cm}\includegraphics[width=7cm]{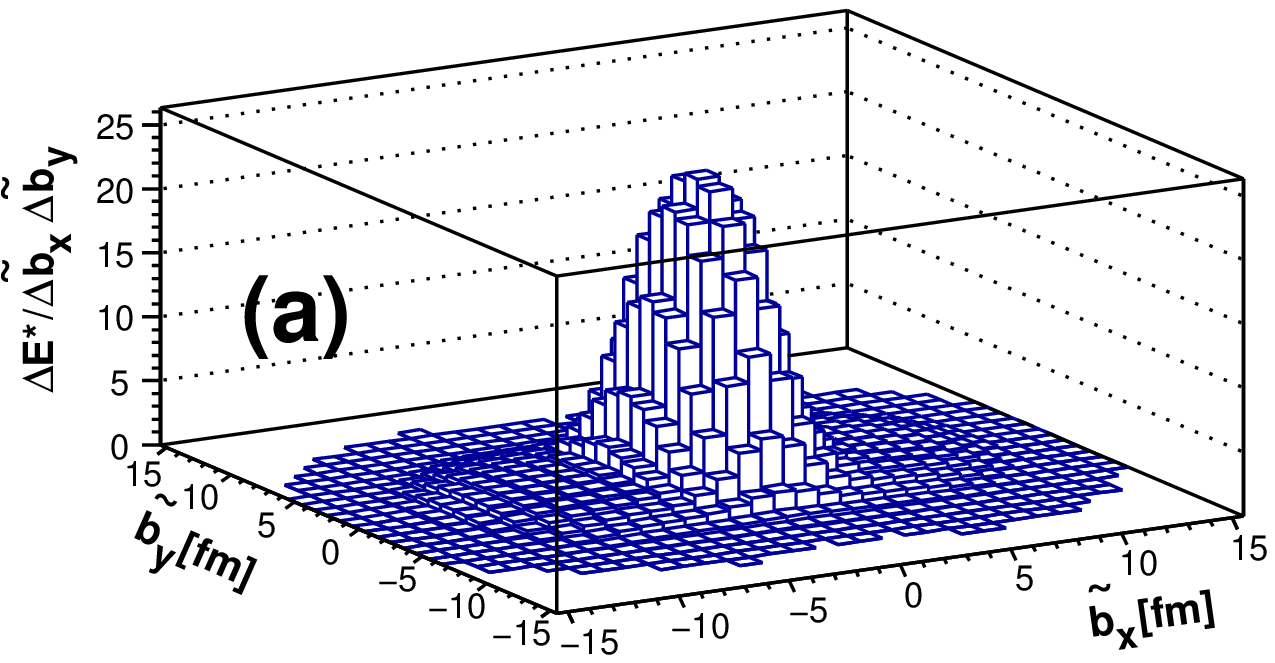}
\includegraphics[width=7cm]{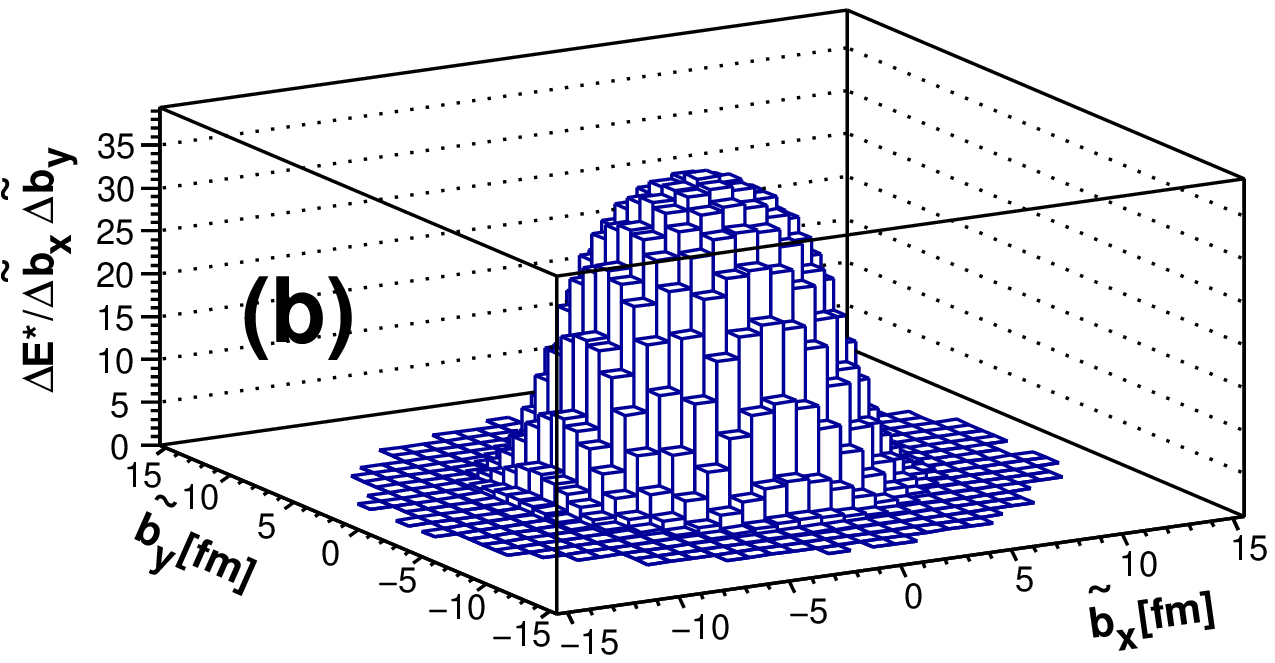}
{\caption
\small 
The $({\tilde b}_x,{\tilde b}_y)$ dependence of the matter excitation energy in 
the bin (strip), for peripheral ($b$=9.72 fm, left panel) and central ($b$=2.55 fm, right panel) Pb+Pb collisions at $\sqrt{s_{NN}}=17.3$~GeV.
The quantity $\Delta E^{*}/\Delta {\tilde b}_x \Delta {\tilde b}_y$ is the strip energy in its own c.m.s. frame, normalized to the size of the  
$({\tilde b}_x,{\tilde b}_y)$ bin.
\label{fig:m_bxby}
}
\end{figure}
 

While, as expected, most of the above trends weaken significantly when going to most central collisions available in the experimental study~\cite{SPS} (0-5\% centrality), it should be underlined that also here it is difficult to speak of ``complete'' stopping of the strips of most probably deconfined partonic matter. A very clear tendency of increase of strip rapidity with increasing ${\tilde b}_x$, from $y\approx$ -2 up to +2 is apparent. The surrounding ``halos'' at $y = \pm y_{beam}$ at the edges of the $({\tilde b}_x,{\tilde b}_y)$ distribution, correspond to regions of very low matter (or mass) density and can be disregarded from our discussion.

Complementary to the above, the distributions of the strip 
excitation energy density (density of the strip energy
in its own c.m.s. frame) in the $({\tilde b}_x,{\tilde b}_y)$ plane of
peripheral and central Pb+Pb collisions are shown in
Fig.~\ref{fig:m_bxby}. As expected, peripheral Pb+Pb collisions at top
SPS energy appear characterized by a relatively narrow region of high
excitation energy, with maximum densities reaching up to 25
GeV/fm$^\mathrm{2}$ in the $({\tilde b}_x,{\tilde b}_y)$ plane. The two
spectator regions where the strip 
energy is equal to its rest mass are clearly visible for higher values of $|{\tilde b}_x|$. In contrast, in central collisions the higher degree of matter ``stopping'' seen in Fig.~\ref{fig:y_bxby} corresponds, in Fig.~\ref{fig:m_bxby}, to a broader ``hot'' region with higher overall excitation energies, reaching up to 35 GeV/fm$^\mathrm{2}$ in the centre of the $({\tilde b}_x,{\tilde b}_y)$ plane.


\section{From QGP matter to hadrons}
\label{qgphadrons}

In our model, after the collision each ``strip'' 
 has its characteristic rapidity $y_S$.
The rapidity distribution of the excited (QGP) matter contained in these strips for a given impact
parameter of the collision $b$ can be written somewhat formally as:
\begin{equation}
\frac{d \sigma}{d y_S}(b) = \int \int \left( \frac{d \sigma}
{d y_S \; d {\tilde b}_x d {\tilde b}_y} ({\tilde b}_x, {\tilde b}_y; b) \right)
d {\tilde b}_x d {\tilde b}_y \; ,
\label{integrated_dsig_dy}
\end{equation} 
where the integrand is differential for a given strip centered
in the transverse impact parameter space around 
$({\tilde b}_x,{\tilde b}_y)$.\\
In practical calculations this can be approximated as
\begin{equation}
\frac{d \sigma}{d y_S}(b) \approx \sum_{i,j} \frac{d \sigma^{ij}(b)}{d y_S}
\; .
\label{on_a_grid}
\end{equation}
where $i$, $j$ run along the finite 
$({\tilde b}_x$ and ${\tilde b}_y)$
bins in the perpendicular plane as shown in Figs~\ref{fig:y_bxby} and~\ref{fig:m_bxby}.

In order to obtain rapidity distributions of resulting particles (pions, kaons)
we have tried the following convolution formula:
%
%
\begin{equation}
\frac{d n}{d y}(b) = \int \int \left( \frac{d n}
{d y d {\tilde b}_x d {\tilde b}_y} ({\tilde b}_x, {\tilde b}_y; b) \right)
d {\tilde b}_x d {\tilde b}_y \; ,
\label{integrated_dn_dy}
\end{equation} 
where the rapidity distribution of pions emerging from the hadronization of one strip located at $({\tilde b}_x,{\tilde b}_y)$ takes the form:
\begin{equation}
\frac{d n}
{d y d {\tilde b}_x d {\tilde b}_y} ({\tilde b}_x, {\tilde b}_y; b) 
= 
\frac{d E_{PROD}^*({\tilde b}_x,{\tilde b}_y ; b)}
{d {\tilde b}_x d {\tilde b}_y}
\;
F( ~ y - y_{S}({\tilde b}_x,{\tilde b}_y ; b )~ ) \; .
\label{one_strip_dn_dy}
\end{equation}
where 
$\frac{d E_{PROD}^*({\tilde b}_x,{\tilde b}_y ; b)}
{d {\tilde b}_x d {\tilde b}_y}$
is the density of the total energy available for particle production in the c.m.s. frame at a given $({\tilde b}_x,{\tilde b}_y)$ point, that is, of its excitation energy minus the rest mass of the two components of the strip coming from nucleus A and B,
\begin{equation}
\frac{d E_{PROD}^*({\tilde b}_x,{\tilde b}_y ; b)}
{d {\tilde b}_x d {\tilde b}_y}
=
\frac{d E^*({\tilde b}_x,{\tilde b}_y ; b)}
{d {\tilde b}_x d {\tilde b}_y}
 -
\frac{dM_A({\tilde b}_x,{\tilde b}_y ; b)}
{d {\tilde b}_x d {\tilde b}_y}
-
\frac{d M_B({\tilde b}_x,{\tilde b}_y ; b)} 
{d {\tilde b}_x d {\tilde b}_y}
\label{de_prod}
\end{equation}
following the discussion we made in section~\ref{simplemodel}. The function 
$F( ~ y - y_{S}({\tilde b}_x,{\tilde b}_y ; b )~ )$ gives the normalized rapidity distribution of pions originating from a given strip of excited matter with rapidity $y_S$. 
In our calculations 
presented here
we tried the following rather flexible 
form~\footnote{We have also tried a standard Gaussian function which was
not sufficient to describe the experimental rapidity distributions given in~\cite{SPS}.}
\begin{equation}
F(\xi) = 
A \exp \left( - \frac{(\sqrt{\xi^{\;2}+\epsilon^{\;2}}~)^{^n}}{n \sigma_y^{\;n}} \right)
\simeq 
A \exp \left( - \frac{|\xi|^n}{n \sigma_y^{\;n}} \right)  \; ,
\label{F}
\end{equation}
where 
$\epsilon$ is a small number which ensures the continuity of derivatives (we use $\epsilon=0.01$),
$\xi = y - y_S$ is the pion rapidity in the strip c.m.s. frame, and the real constants $A$, $\sigma_y$, and $n$ are free parameters. 
These three parameters 
will be tuned to fit the experimental data, but we assume\footnote{The detailed validity of this assumption will be tested in section~\ref{negative_pion}.} their full independence on ${\tilde b}_x$, ${\tilde b}_y$, and $b$, that is, on strip position and collision centrality.

We note that the transition between excited (initially QGP) matter and produced particles is 
then entirely contained in the overall constant $A$ and the above universal
(impact parameter independent) modified exponential function.

Several features of the simple hadronization scheme proposed in Eqs.~\ref{integrated_dn_dy}-\ref{F} are noteworthy:

\begin{itemize}
\item[$\bullet$]
 For each considered $({\tilde b}_x$, ${\tilde b}_y)$ strip of excited matter, 
the dependence of 
its hadronization into pions 
on strip position and 
collision
centrality is uniquely defined by the dependence of the strip rapidity, $y_S$, and of its (excitation energy - rest mass), 
$\frac{d E_{PROD}^*}
{d {\tilde b}_x d {\tilde b}_y}$, respectively on ${\tilde b}_x$, ${\tilde b}_y$, and $b$. 
%
Both $y_S$ and $\frac{d E_{PROD}^*}{d {\tilde b}_x d {\tilde b}_y}$
are directly given by energy-momentum conservation, as discussed in section~\ref{simplemodel} and then shown 
in Figs~\ref{fig:y_bxby} and~\ref{fig:m_bxby}.
\item[$\bullet$]
 Specifically, the shape of the produced pion rapidity distribution in the strip c.m.s. frame, given by Eq.~\ref{F}, remains independent of the strip position and collision centrality.
\item[$\bullet$]
Statistically speaking, our hadronization scheme of excited matter into pions preserves energy conservation as the number of produced pions is directly proportional to the total strip energy available for particle production, 
explicit from Eq.~\ref{one_strip_dn_dy}.
\end{itemize}

In practical calculations we have to use finite 
($\Delta {\tilde b}_x, \Delta {\tilde b}_y$) 
bins, instead of 
($d {\tilde b}_x, d {\tilde b}_y$) as addressed above. Analogically to 
Eqs.~\ref{integrated_dn_dy}-\ref{F},
the approximate formula for the emitted pion rapidity distribution takes then the somewhat simplified form: 
\begin{equation}
\frac{d n}{d y}(b) \approx \sum_{i,j} 
\left(\frac{\Delta E_{PROD}^*({\tilde b}_x,{\tilde b}_y ; b)}
{\Delta {\tilde b}_x \Delta {\tilde b}_y}\right)
\;
{\Delta {\tilde b}_x \Delta {\tilde b}_y}
\;
\;
F( ~ y - y_{S}({\tilde b}_x,{\tilde b}_y ; b )~ ) \; ,
\label{integrated_dn_dy_approx}
\end{equation} 
where
\begin{equation}
\frac{\Delta E_{PROD}^*({\tilde b}_x,{\tilde b}_y ; b)}
{\Delta {\tilde b}_x \Delta {\tilde b}_y}
=
\frac{\Delta E^*({\tilde b}_x,{\tilde b}_y ; b)}
{\Delta {\tilde b}_x \Delta {\tilde b}_y}
 -
\frac{\Delta M_A({\tilde b}_x,{\tilde b}_y ; b)}
{\Delta {\tilde b}_x \Delta  {\tilde b}_y}
-
\frac{\Delta  M_B({\tilde b}_x,{\tilde b}_y ; b)} 
{\Delta  {\tilde b}_x \Delta  {\tilde b}_y} \; .
\label{deltae_prod}
\end{equation}
Here, the distributions of $y_{S}({\tilde b}_x,{\tilde b}_y ; b )$ and 
$\frac{\Delta E_{PROD}^*({\tilde b}_x,{\tilde b}_y ; b)}
{\Delta {\tilde b}_x \Delta {\tilde b}_y}$
have been presented in Figs~\ref{fig:y_bxby} and~\ref{fig:m_bxby} 
for the case of peripheral and central collisions at top SPS energy, 
$\frac{\Delta M_{A~(B)}({\tilde b}_x,{\tilde b}_y ; b)}
{\Delta {\tilde b}_x \Delta  {\tilde b}_y}$ are the ``cold'' mass densities of the two components of the strip coming from nucleus A and B, and the one-strip rapidity spectrum $F(\xi)$, Eq.~\ref{F}, can be obtained from a simple Monte Carlo generation.

\section{Negative pion spectra in Pb+Pb collisions}
\label{negative_pion}

The NA49 experiment at the SPS measured rapidity spectra of negative pions in a broad rapidity range for Pb+Pb collisions, specifically also at top SPS energy~\cite{SPS}. The spectra were measured as a function of collision centrality, quantified therein by the average number of wounded nucleons~\cite{wnm} calculated using the VENUS model~\cite{Werner}, account taken of the experimental conditions. Apart from the well-known increase in absolute density, a continuous trend of noticeable narrowing of the shape of the pion $dn/dy$ distribution with increasing centrality of the collision is apparent in the experimental data.

\begin{figure}
\includegraphics[width=5cm]{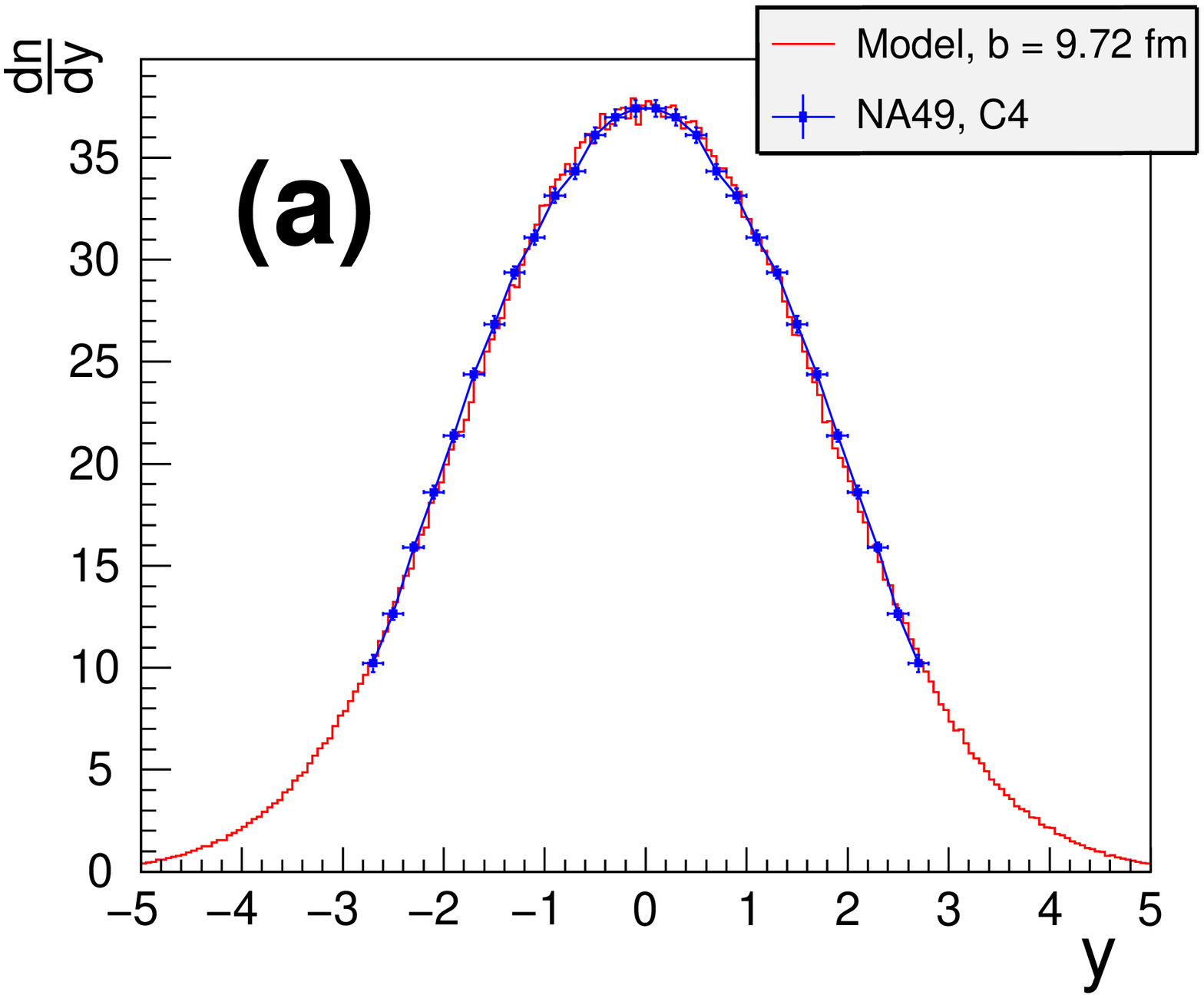}
\includegraphics[width=5cm]{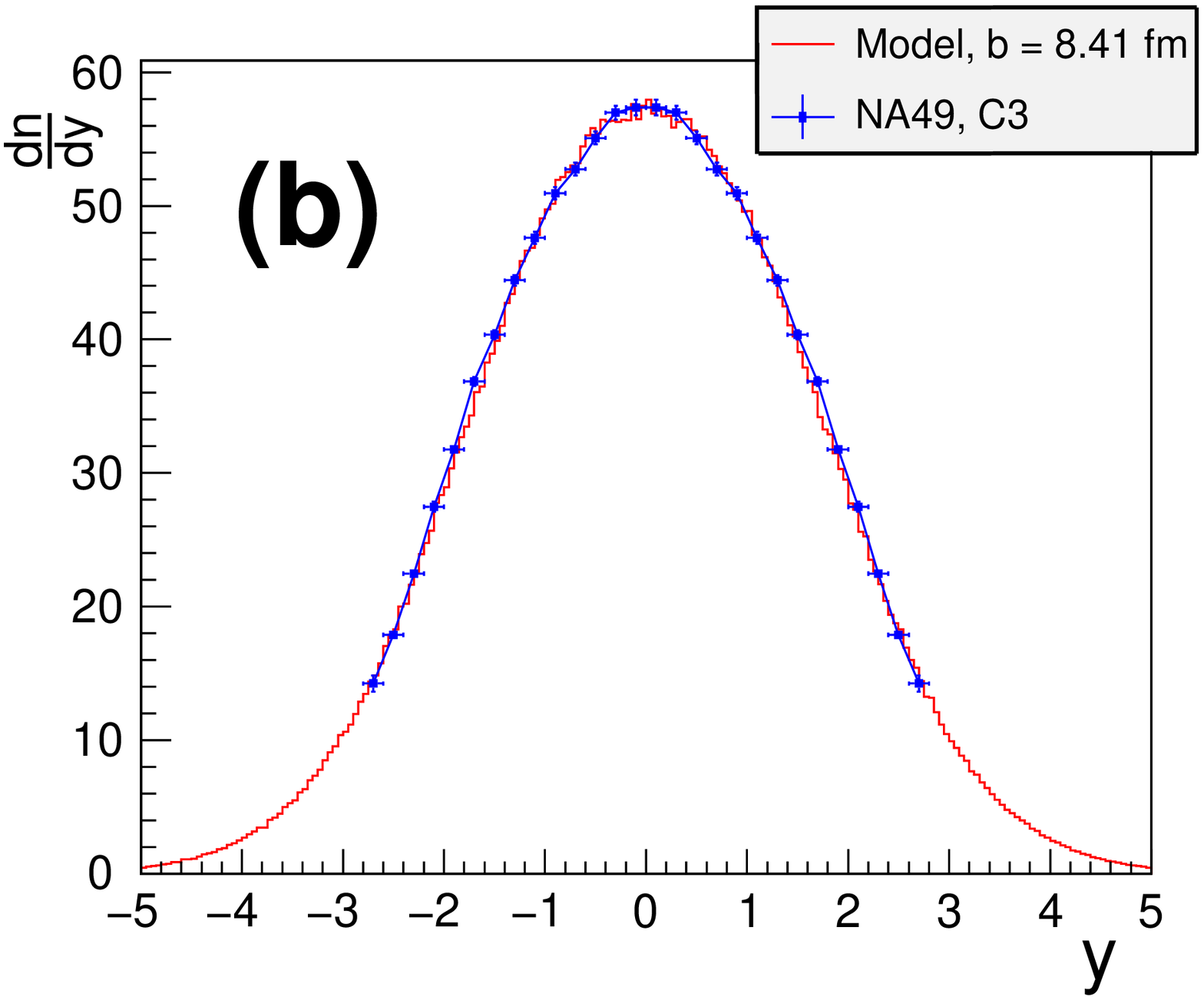}
\includegraphics[width=5cm]{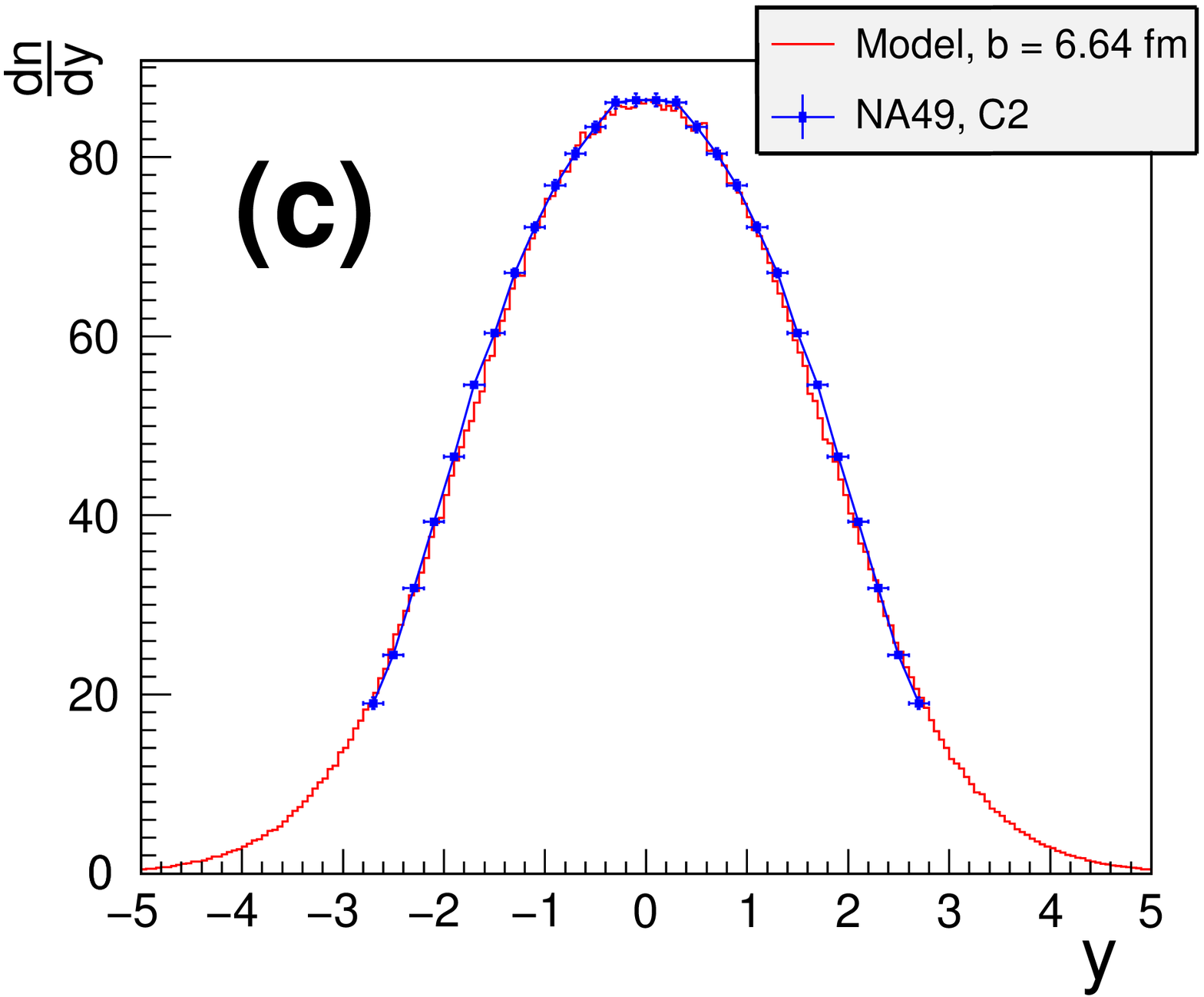}\\
\includegraphics[width=5cm]{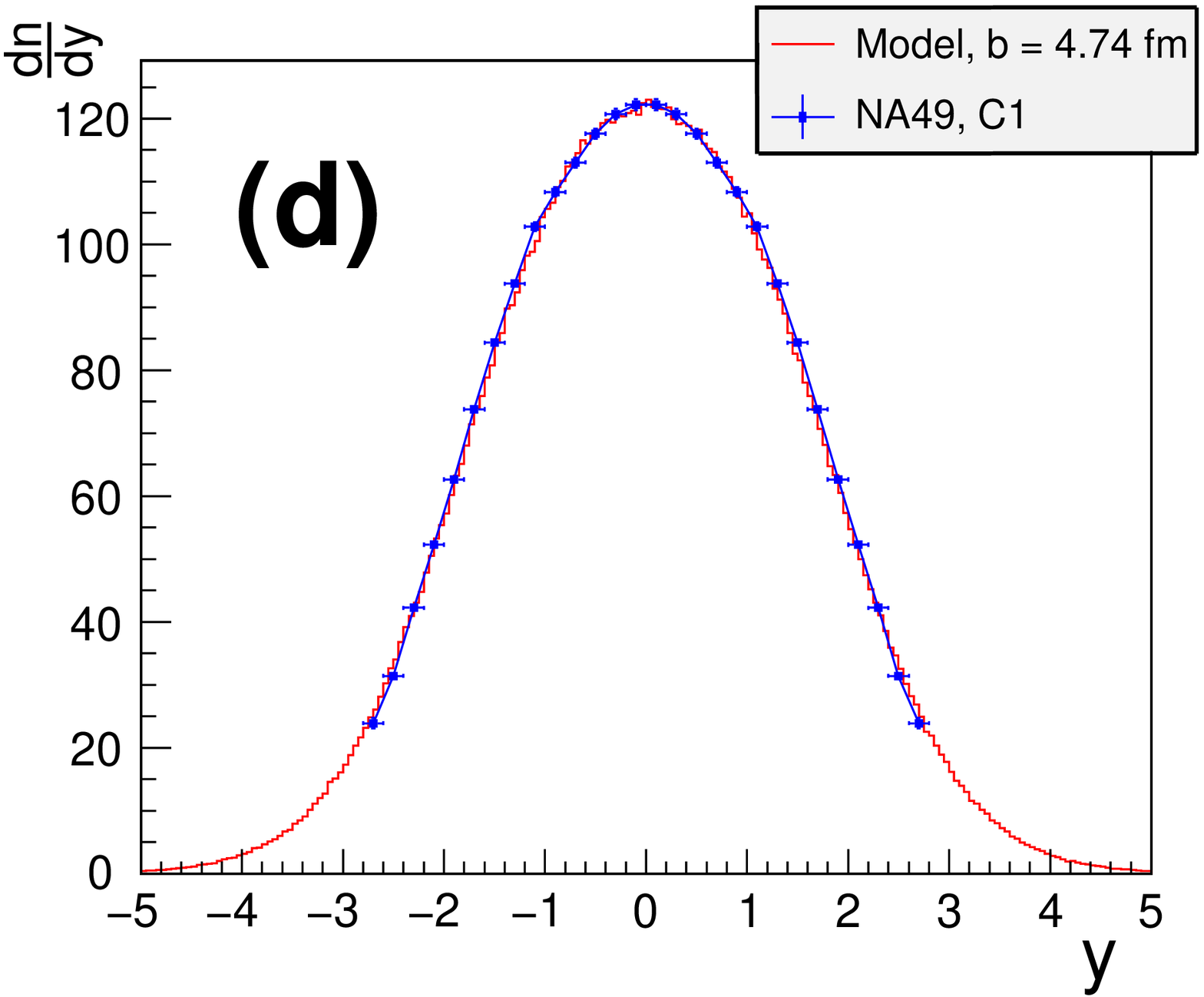}
\includegraphics[width=5cm]{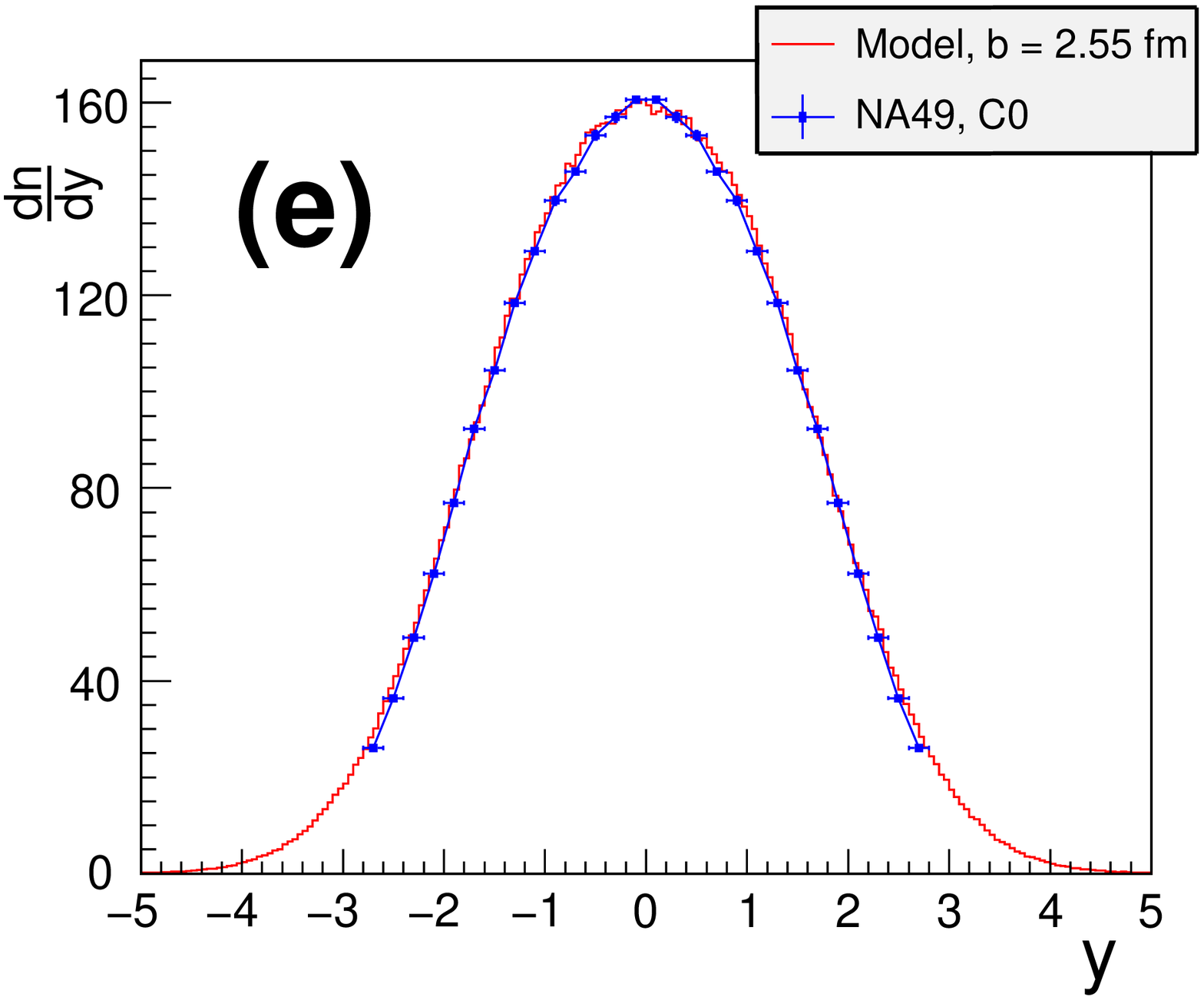}
{\caption
\small Rapidity distributions of negative pions measured by the NA49 experiment~\cite{SPS} in five centrality selected samples of Pb+Pb collisions at $\sqrt{s_{NN}}$=17.3~GeV, put together with our model predictions as described in the text. Here we have taken $\sigma_y$ = 1.475 and $n$ = 2.55.
\label{fig:dsigma_dy_fordifferentb}
}
\end{figure}

This is illustrated in Fig.~\ref{fig:dsigma_dy_fordifferentb} where on top of the five NA49 centrality samples, our model predictions for five different 
impact parameters 
(b = 
9.72, 
8.41,
6.64, 
4.74, 
2.55
fm)  
are shown. As mentioned in section~\ref{pbpb}, a Glauber approach is used to translate the mean number of wounded nucleons published by the NA49 collaboration into the impact parameter. Our results are obtained from a global fit of the two parameters $\sigma_y$ and $n$ from Eq.~\ref{F}, but allow for an additional, {very small} variation of the normalization constant $A$ (we will address this issue separately below).

\begin{figure}[!ht]
\includegraphics[width=6.5cm]{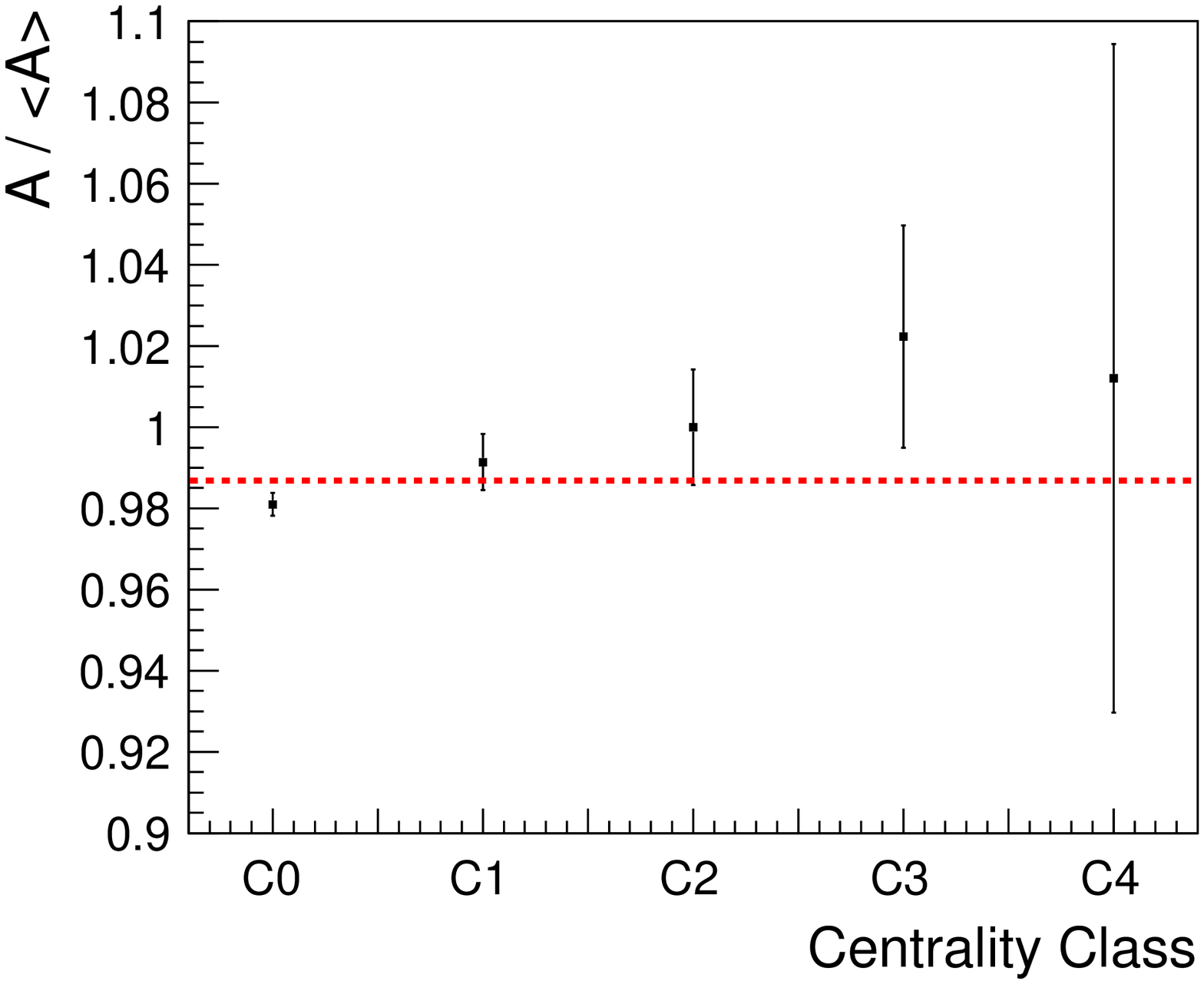}
{\caption
\small Normalization factor $A$ from Eq.~\ref{F}, for the five centrality classes of Pb+Pb collisions
considered in this paper, with respect to its average value. The error bars assumed for the data points are explained in the text.
A tentative horizontal line is drawn in the plot.
\label{fig:normalization_of_centrality_improved}
}
\end{figure}

Evidently, with $\sigma_y$ = 1.475 and $n$ = 2.55, 
our model gives a very good description of the shape of the NA49 rapidity spectra and of its evolution with centrality. The variation of the normalization parameter $A$, Eq.~\ref{F}, as a function of centrality is shown in Fig.~\ref{fig:normalization_of_centrality_improved}. The error bars assumed for $A$ correspond to the same relative uncertainty as that estimated for the number of wounded nucleons published in~\cite{SPS}. Our underlying logic is that an increase (decrease) of the true number of wounded nucleons w.r.t.~to its value postulated in~\cite{SPS} will give a proportional increase (decrease) of the measured $dn/dy$ density and therefore the same increase (decrease) of the fitted value of $A$. 


As it is evident from
Fig.~\ref{fig:normalization_of_centrality_improved}, the absolute
normalization of pion rapidity spectra defined in our 
{new realization of the fire-streak} model by Eq.~\ref{F} as well as Eqs.~\ref{integrated_dn_dy_approx} and~\ref{deltae_prod}, fits the experimental data up to a $\pm$2\% accuracy (to be compared with an increase of absolute $dn/dy$ density of more than 300\% from peripheral to central Pb+Pb collisions apparent in Fig.~\ref{fig:dsigma_dy_fordifferentb}). This is basically comparable to the uncertainty induced by the estimated number of wounded nucleons addressed above. Thus we conclude that our model assumption of full independence of the three free parameters $A$, $\sigma_y$ and $n$ on strip position and centrality (section~\ref{qgphadrons}, Eq.~\ref{F}) describes the experimental data~\cite{SPS} within its present systematic accuracy~\footnote{While this is premature in the view of the accuracy of this analysis, a small systematic variation of $A$ with centrality cannot be excluded. For instance, a small increase towards peripheral collisions possibly suggested by Fig.~\ref{fig:normalization_of_centrality_improved} would correspond to a larger fraction of available energy going into pions w.r.t. to other less abundant particles, like kaons.}. 

We come to the conclusion that as such, our model formulated in sections~\ref{simplemodel} and~\ref{qgphadrons}
gives a complete description of $dn/dy$ distributions of pions in the full considered range of collision centrality. We note that our model does not contain any explicit assumption on wounded nucleons nor wounded nucleon scaling. In other terms, the shape and evolution of these spectra as a function of $b$ can be explained solely as a consequence of local energy conservation for the excited matter intially created in the collision (Figs~\ref{fig:y_bxby}, \ref{fig:m_bxby}), preserved - in the statistical sense - in our somewhat 
simplified
hadronization scheme defined by Eqs~\ref{F}-\ref{deltae_prod}.

\section{Narrowing of rapidity distributions with increasing centrality}
\label{narrowing}

\begin{figure}[!ht]
\hspace*{-0.5cm}\includegraphics[width=7.5cm]{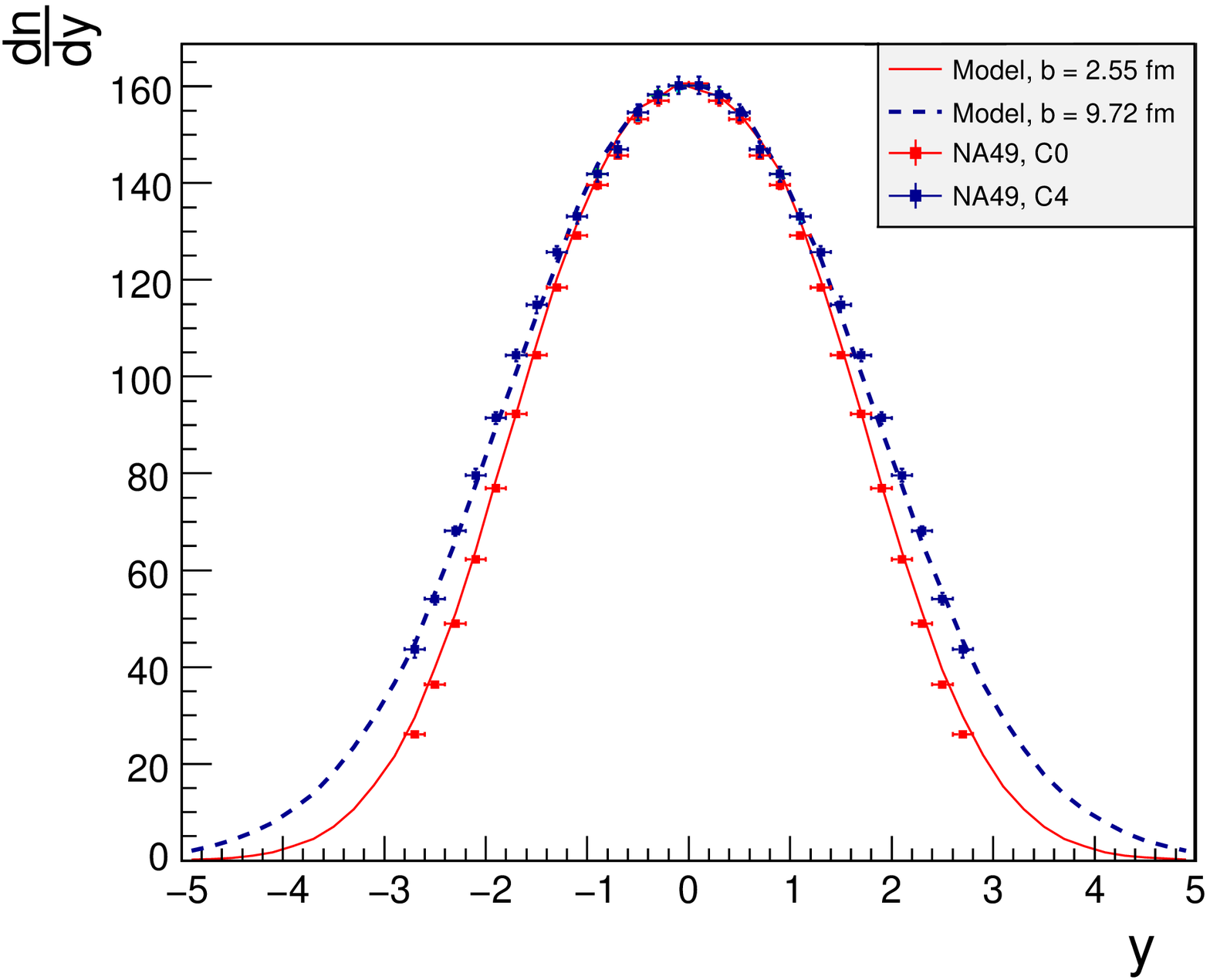}
{\caption
\small 
Shape comparison for $\pi^-$ rapidity spectra in peripheral (``C4'') and central (``C0'') collisions obtained by the NA49 experiment~\cite{SPS}, put together with our model predictions. For peripheral collisions, both experimental data and model 
results
from Fig.~\ref{fig:dsigma_dy_fordifferentb} are multiplied up by an arbitrary factor in order to match the distributions in central collisions at $y=0$.
\label{fig:narrowing}
}
\end{figure}

We will now specifically focus on the centrality dependence of the shape of pion rapidity spectra. Fig.~\ref{fig:narrowing} shows the shape comparison of experimental data in central and peripheral Pb+Pb collisions, superimposed with the prediction of our model. A narrowing of the distribution with increasing centrality is evident in the NA49 data and very well reproduced by our model curves. 

We note that in our model, the above narrowing trend is a 
natural effect, resulting exclusively from realistic collision geometry and energy-momentum conservation. The narrowing of $dn/dy$ spectra of pions with decreasing impact parameter is a direct reflection of the different degrees of stopping of the initially created matter as shown in Fig.~\ref{fig:y_bxby} for peripheral and central collisions. The subsequent hadronization process brings no further centrality dependence to the shape of the pion rapidity distribution, as discussed in section~\ref{qgphadrons}. We conclude that indeed, the effect of narrowing of rapidity spectra is to be regarded as resulting from four-momentum conservation rather than from any sophisticated dynamics.

\begin{figure}[!hb]
\hspace*{-0.6cm}\includegraphics[width=7.4cm]{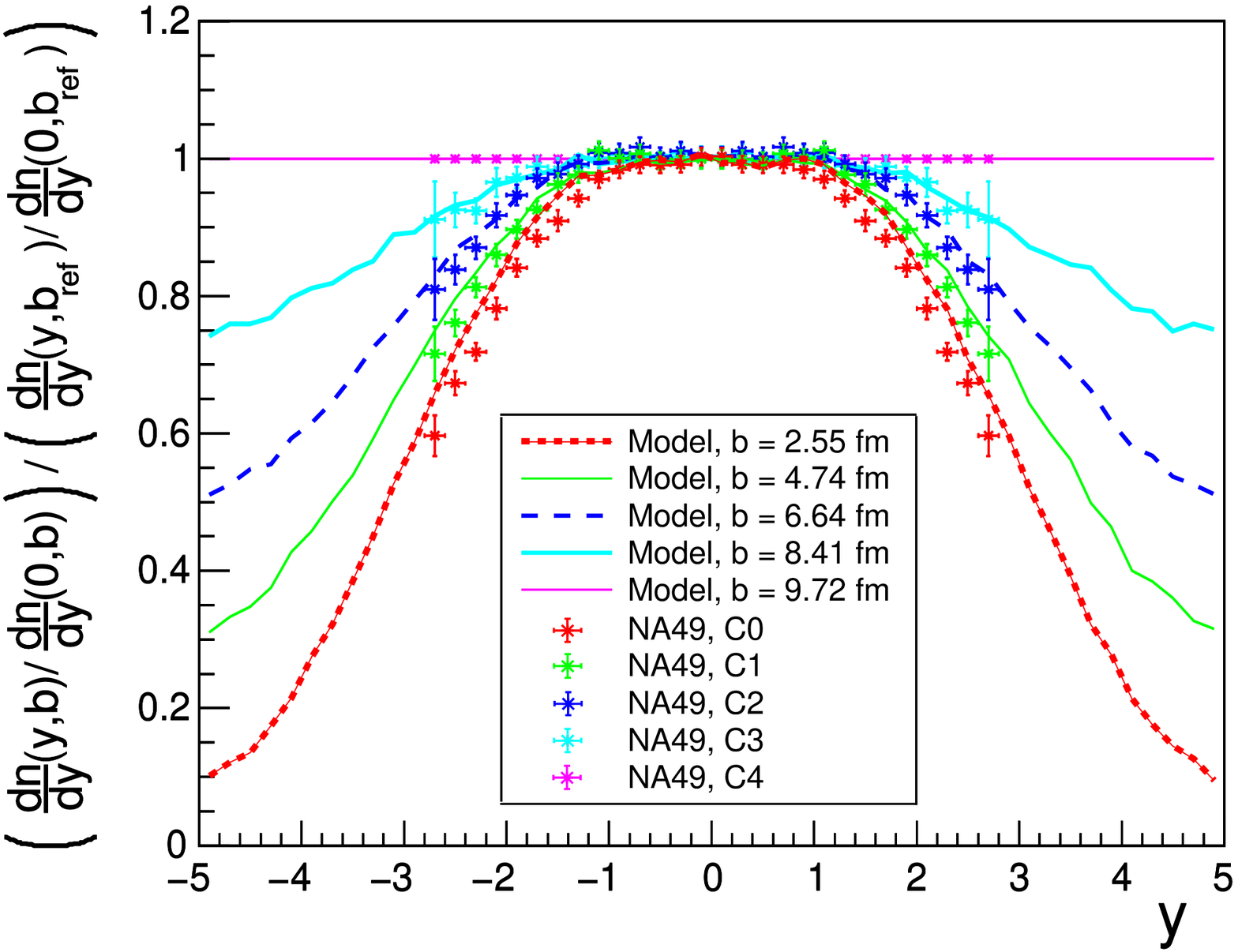}
{\caption
\small Shapes of rapidity spectra, normalized to most
peripheral Pb+Pb collisions. The experimental data points are obtained from the NA49 data~\cite{SPS} by adding the errors in quadrature, and compared to our model predictions.
\label{fig:shape_description}
}
\end{figure}

A further, more detailed study of this effect can be accomplished by the use of a specific quantity:
\begin{equation}
R_{shape}(y ; b , b_{ref})=
\left(
\frac{dn}{dy}(y,b)/\frac{dn}{dy}(0,b)
\right) {\Big/}
\left(
\frac{dn}{dy}(y,b_{ref})/\frac{dn}{dy}(0,b_{ref})
\right) 
\label{Rshape}
\end{equation} 
which gives the relative shape modification of the pion $dn/dy$ distribution at a given collision impact parameter $b$, w.r.t. the same distribution at a ``reference'' impact parameter $b_{ref}$. Here we take $b_{ref}=9.72$~fm which normalizes $R_{shape}(y ; b , b_{ref})$ to the most peripheral sample of Pb+Pb collisions considered in Fig.~\ref{fig:dsigma_dy_fordifferentb}.
Our result on $R_{shape}$ is presented in Fig.~\ref{fig:shape_description}, together with the corresponding quantity extracted from the NA49 data where the ``C4'' sample is used to normalize the other experimental distributions.
The narrowing of the experimental pion distribution appears as a smooth trend as a function of centrality. In a natural way, this feature is reproduced by our model as a pure result of smoothly changing collision geometry put together with local energy-momentum conservation, and {with no extra dynamics}. We state that quantitatively, our model reproduces very well the data (it describes $\sim$90\% of the effect for the {worst} case of central collisions). We leave for future studies the question whether the remaining small discrepancy is to be attributed to additional experimental 
uncertainties~\footnote{We
 note that apart from the statistical uncertainties drawn in Fig.~\ref{fig:shape_description}, an overall systematic error of 10\% (5\%) is quoted for pion spectra in peripheral (all other) Pb+Pb data samples discussed in Ref.~\cite{SPS}.}, 
our approximations,
or possible dynamical effects. On the other hand, even 
if the extrapolation of our results up to high rapidity
in Fig.~\ref{fig:shape_description} should be taken with some caution, we note the interest of possible new measurements of fast pion production for a further clarification of the role of energy-momentum conservation
 in heavy ion collisions.

\section{Summary and concluding remarks}
\label{conclusions}

We will now try to conclude what we consider the main lessons to be drawn from our study.

\begin{itemize}
\item[$\bullet$]
We proposed a very simple {model, which can be regarded as a new realization of the nuclear fire-streak concept \cite{M1978,GKW1978}. This model was} uniquely based on a proper description of collision geometry originating from realistic density distributions of the two incoming nuclei, and on rigorous local energy-momentum conservation for the initially created matter in the transverse $({\tilde b}_x,{\tilde b}_y)$ 
 plane, followed by a simple hadronization scheme. The latter scheme assumed that the number of pions produced from a single $({\tilde b}_x,{\tilde b}_y)$ strip was proportional to its total energy available for particle production and, in its own c.m.s. frame, independent of ${\tilde b}_x$, ${\tilde b}_y$,  and collision centrality.
\item[$\bullet$]
 This model describes the whole centrality dependence of negative pion $dn/dy$ spectra measured by the NA49 experiment in Pb+Pb collisions at $\sqrt{s_{NN}}$=17.3~GeV, both in terms of absolute $dn/dy$ yields as 
well as
of the narrowing of rapidity spectra from peripheral to central
collisions. This latter trend is in our 
{model} 
a pure consequence of energy and momentum conservation. 
\item[$\bullet$]
 In order to provide this description, it 
was 
sufficient to use 
 three centrality-independent parameters ($A$, $n$ and $\sigma_y$) which define  the (unique) shape of the single-strip pion spectrum. This means that once adjusted to a single collision impact parameter $b_{ref}$, the whole centrality dependence of the longitudinal evolution of pion production could be correctly deduced from the conservation of energy-momentum.
\item[$\bullet$]
 We noted only very small discrepancies in our comparison between the experimental data and the model. For the time being we disregard 
them
as they can be blamed on the NA49 systematic errors quoted in~\cite{SPS}. More precise measurements would be needed in order to 
understand
whether more refined corrections should be applied to the simple picture emerging 
from our study. 
\end{itemize}

From the above we conclude that the longitudinal evolution of the system of hot and dense matter initially created in the heavy ion collision at SPS energies is dominantly defined 
purely by 
energy-momentum conservation. A rigourous treatment of local four-momentum conservation as included in our model appears sufficient to give a basic description of pion production as a function of centrality and rapidity. This does obviously not imply that our simplistic approach presented here is to be taken {\em \`{a} la lettre}. Dynamical factors like for instance, the building up of collective flow phenomena will, as it is widely known~\cite{transverse_expansion,azimuthal_anisotropies}, impose a transverse expansion and azimuthal anisotropies in the process of particle emission, and the whole system will undergo a phase transition which will affect a number of observables. It is however encouraging that at least for the case of pion production at top SPS energy considered here, our fairly simple treatment can provide insight into its longitudinal evolution, rather poorly known 
for theoretical and partially 
experimental reasons.

On the other hand, we wish to comment on the space-time picture of the
Pb+Pb collision which emerges from our study. We did not adress the
issue of the physical nature of the idealized $({\tilde b}_x,{\tilde
  b}_y)$ ``strips'' ({fire-streaks}) considered in this model which we presume to consist, depending on their excitation energy, of deconfined partonic or excited hadronic matter. It seems however clear that quite independently of its exact nature, the distribution of bulk of this matter in configuration $(x,y,z)$ space will also carry the imprint of its evolution in rapidity (see Ref.~\cite{bialas} for comparison). In that respect the following remarks are in place:
\begin{itemize}
\item[$\bullet$]
At least in non-central collisions, 
``participant'' and ``spectator'' 
systems may be difficult to differentiate exactly. 
Local energy-momentum conservation implies the presence of ``streams'' of excited matter moving close to spectator rapidity as presented in Fig.~\ref{fig:ideowy_po}. This corresponds to ``sources'' of pion emission placed at a very small $(x,y,z)$ distance from the 
``cold'' spectator matter moving close to beam rapidity. 
%
%
\item[$\bullet$]
As we expect the ``strips'' ({fire-streaks}) of created matter to be characterized by some (average) proper hadronization, or freeze-out time $\tau_f$, it seems natural to expect that relativistic effects will impose a sequential ordering in the collision c.m.s. 
time moment
of pion formation, $t_f$, as a function of 
rapidity. Consequently, the 
longitudinal distance $d_E$ between the pion emission point at freeze-out and the corresponding position of the {nearest} spectator system,
\begin{equation}
d_E = z_{spec}(y_{spec},t_f) -  z_\pi(y,t_f) \;
\label{de}
\end{equation}
 as illustrated in Fig.~\ref{fig:ideowy_po}, will strongly decrease as a function of 
strip rapidity. This is consistent with our preliminary results on the rapidity dependence of $d_E$ as deduced from spectator-induced electromagnetic (EM) interactions, see Ref.~\cite{wpcf}.
\item[$\bullet$]
The presence of pion emission close to spectator matter as addressed above opens the way to reinteraction processes induced either by the strong or the EM force. Such reinteraction phenomena could be responsible for a strong enhancement of pion production at $y\approx y_{beam}$, low $p_T$ which is visible in preliminary NA49 data on peripheral Pb+Pb collisions w.r.t. proton-proton reactions, see Ref.~\cite{Rybicki2015}.
%
\item[$\bullet$]
Finally, while our model in principle provides a very efficient mechanism for ``stopping''~\cite{busza84} of baryonic matter as a function of centrality, we underline that 
a
caution in drawing strong conclusions on this basis seems 
recommended
in this case. One should remember that the baryonic content of the ``strips'' as a function of $z$ will be defined by dynamical processes connected, for instance, to the original quark-gluon distributions in the intially created matter. As such, we do not believe that these can be approximated by our simplified Eq.~\ref{F}.
\end{itemize}

{To sum up,
 our model work presented in this paper allows to understand 
the dependence of rapidity distributions, in particular their
widths, on impact parameter (centrality)
of the heavy ion collision in the SPS energy regime.
The main aim of this work was to provide a simple, yet realistic, model which can
be used to address quantitatively electromagnetic effects
generated by spectators. In our previous works on EM effects, cited in
this paper, we found that fast pions are produced closer to
spectators than slow ones. This observation was confirmed in our studies
of charged pion directed flow $v_1$ 
\cite{Rybicki_v1}.

As discussed 
above,
our model qualitatively explains 
the
observations
we
made in our previous studies of EM effects due to spectators.
A next obvious step will be combining of the 
version 
of the fire-streak model 
proposed by us 
together with electromagnetic effects.
The observables to be explained are final state single charged pion
distributions in rapidity and/or transverse momentum,
$\pi^+/\pi^-$ ratios as well as the EM splitting of directed flow 
for 
charged pions
\cite{wpcf,Rybicki2015,Rybicki_meson2016}.
%
We wish to add here, that the fire-streak concept in its version proposed here can not only explain the small electromagnetic splittings of pion $v_1$ but also has the
potential to explain the bulk effect for directed flow, i.e. its departure
from zero.
Equiped with such a simple, ready to use apparatus we will continue our
studies of EM effects.}\\

{\bf Acknowledgement}\\
{We thank our referee for pointing to us the similarity of our approach to the nuclear fire-streak concept.}
This work was supported by the National Science Centre, Poland
(grant no. 2014/14/E/ST2/00018).






\begin{thebibliography}{99}

\bibitem{twospec07}   
 A. Rybicki, A. Szczurek, Phys.\ Rev.\ C {\bf 75}, 054903 (2007).

\bibitem{Rybicki_v1} 
  A.~Rybicki and A.~Szczurek,
  Phys.\ Rev.\ C {\bf 87}, 
  054909 (2013).

\bibitem{twospec_auau} 
  A.~Rybicki and A.~Szczurek,
  arXiv:1405.6860 [nucl-th].

\bibitem{wpcf} 
  A.~Rybicki
  {\it et al.},
  Acta Phys.\ Polon.\ Supp.\  {\bf 9}, 303 (2016).
  
\bibitem{M1978}
W.D. Myers, 
Nucl. Phys. {\bf A296} (1978) 177.

\bibitem{GKW1978}
J. Gosset, J.I. Kapusta and G.D. Westfall,
Phys. Rev. {\bf C18} (1978) 844.

\bibitem{MCS2001}
V.K. Magas, L.P. Csernai and D.D. Strottman,
Phys. Rev. {\bf C64} (2001) 014901.

\bibitem{MCS2002}
V.K. Magas, L.P. Csernai and D.D. Strottman,
Nucl. Phys. {\bf A712} (2002) 167.

\bibitem{MK2002}
I.N. Mishustin and J.I. Kapusta,
Phys. Rev. Lett. {\bf 88} (2002) 112501.

\bibitem{Trzcinska01}
A. Trzci\'nska {\it et al.}, Phys. Rev. Lett. {\bf 87}, 082501 (2011).

\bibitem{atph} 
 A.Trzci\'nska, Ph.D. Thesis, Heavy Ion Laboratory, Warsaw University, May 
2001; 
({\em in Polish}).

\bibitem{SPS} 
  T.~Anticic {\it et al.},
  Phys.\ Rev.\ C {\bf 86}, 054903 (2012).

\bibitem{Glauber} 
  R.~J.~Glauber,
  Phys.\ Rev.\  {\bf 100}, 242 (1955).

\bibitem{wnm}
A.~Bia\l{}as, M.~Bleszy\'nski and W.~Czy\.z,
Nucl.\ Phys.\ B {\bf 111}, 461 (1976).

\bibitem{Werner} 
  K.~Werner,
  Phys.\ Rept.\  {\bf 232}, 87 (1993).

\bibitem{transverse_expansion}
  W.~Broniowski and W.~Florkowski,
  Phys.\ Rev.\ Lett.\  {\bf 87}, 272302 (2001).

\bibitem{azimuthal_anisotropies}
  J.~Y.~Ollitrault,
  Phys.\ Rev.\ D {\bf 46}, 229 (1992).



\bibitem{bialas} 
  A.~Bia\l{}as, A.~Bzdak and V.~Koch,
  arXiv:1608.07041 [hep-ph].

\bibitem{Rybicki2015} 
  A.~Rybicki, A.~Szczurek and M.~K\l{}usek-Gawenda,
  Acta Phys.\ Polon.\ B {\bf 46}, no. 3, 737 (2015).

\bibitem{busza84}
  W.~Busza and A.~S.~Goldhaber,
  Phys.\ Lett.\  {\bf 139B}, 235 (1984).


\bibitem{Rybicki_meson2016} 
 {A.~Rybicki {\it et al.}, 
  EPJ Web Conf.\  {\bf 130}, 05016 (2016).}


\end{thebibliography}
\end{document}